\documentclass[a4paper,11pt]{article}
\pdfoutput=1
\usepackage[normalem]{ulem}

\usepackage{jcappub} 
                    
\usepackage{amsmath}
\usepackage{amssymb}
\usepackage{color}
\usepackage{graphicx}
\usepackage{comment}
\usepackage{multirow}
\usepackage[dvipsnames]{xcolor}
\usepackage[utf8]{inputenc}
\usepackage{caption}
\usepackage{epstopdf}

\usepackage[T1]{fontenc} 

\include{definition}

\newcommand{\HI}{\rm H~{\sc i}}
\newcommand{\HII}{\rm H~{\sc ii}}

\newcommand{\TB}{\delta T_{\rm b}}
\newcommand{\MSUN}{{\rm M}_{\odot}}

\newcommand{\TS}{T_{\rm S}}
\newcommand{\TK}{T_{\rm K}}
\newcommand{\TCMB}{T_{\gamma}}
\newcommand{\TRAD}{T_{\rm rad}}

\newcommand{\lya}{\rm {Ly{\alpha}}}
\newcommand{\lyb}{\rm {Ly{\beta}}}
\newcommand{\OmegaB}{\Omega_{\rm B}}
\newcommand{\Omegam}{\Omega_{\rm m}}

\newcommand{\Tvir}{T_{\textup{vir}}}

\title{Astrophysical information from the Rayleigh-Jeans Tail of the CMB}

\author[a,b]{Raghunath Ghara,}
\author[c]{Garrelt Mellema,}
\author[d,b,e]{Saleem Zaroubi}

\affiliation[a]{ARCO (Astrophysics Research Center), The Open University of Israel, 1 University Road, PO Box 808, Ra'anana 4353701, Israel}
\affiliation[b]{Department of Physics, Technion, Haifa 32000, Israel}
\affiliation[c]{Department of Astronomy \& Oskar Klein Centre, Stockholm University, AlbaNova, SE-10691 Stockholm, Sweden}
\affiliation[d]{Department of Natural Sciences, The Open University of Israel, 1 University Road, PO Box 808, Ra'anana 4353701, Israel}
\affiliation[e]{Kapteyn Astronomical Institute, University of Groningen, PO Box 800, 9700AV Groningen, the Netherlands}

\emailAdd{ghara.raghunath@gmail.com}
\emailAdd{garrelt.mellema@astro.su.se}

\abstract{One of the explanations for the recent EDGES-LOW band 21-cm measurements of a strong absorption signal around 80~MHz is the presence of an excess radio background to the Cosmic Microwave Background (CMB). Such excess can be produced by the decay of unstable particles into small mass dark photons which have a non-zero mixing angle with electromagnetism. We use the EDGES-LOW band measurements to derive joint constraints on the properties of the early galaxies and the parameters of such a particle physics model for the excess radio background. A Bayesian analysis shows that a high star formation efficiency and X-ray emission of $4-7 \times 10^{48} ~\rm erg$ per solar mass in stars are required along with a suppression of star formation in halos with virial temperatures $\lesssim 2\times 10^4$ K. The same analysis also suggests a 68 percent credible intervals for the mass of the decaying dark matter particles, it's lifetime, dark photon mass and the mixing angle of the dark and ordinary photon oscillation of $[10^{-3.5}, 10^{-2.4}]$ eV, $[10^{1.1}, 10^{2.7}]\times 13.8 ~\rm Gyr$, $[10^{-12.2}, 10^{-10}]$ eV and $[10^{-7}, 10^{-5.6}]$ respectively. This implies an excess radio background which is $\approx 5.7$ times stronger than the CMB around 80~MHz. This value is a factor $\sim 3$ higher than the previous predictions which used a simplified model for the 21-cm signal. }

\begin{document}
\maketitle
\flushbottom

\section{Introduction}
\label{sec:intro}
Understanding the properties of the very first sources of light that formed in the Universe is one of the goals of today's astronomy. Direct observations of those early sources such as the first stars or high-mass X-ray binaries (HMXBs) is currently impossible. However, it has been shown that the redshifted 21-cm signal from the neutral hydrogen from the epoch when first sources formed in the Universe (often called the `Cosmic Dawn') contains substantial information about the early source population \cite[see e.g.][for reviews]{Pritchard12, 2013ASSL..396...45Z}.  Thus, observations of this signal constitute the most promising indirect probe of the very first sources of radiation that formed in the Universe. 

Motivated by this, many radio telescope systems have been designed to attempt to measure the signal. These systems fall into two categories. The first one consists of large radio interferometers which are capable to measure the redshift evolution of the spatial fluctuations of the signal in terms of statistical quantities such as power-spectrum. Radio telescopes systems such as the Low Frequency Array (LOFAR)\footnote{http://www.lofar.org/} \citep{2020MNRAS.493.1662M, 2022MNRAS.509.3693M}, the Giant Metrewave Radio Telescope (GMRT)\footnote{http://www.gmrt.tifr.res.in}\citep{ghosh12, paciga13}, the Precision Array for Probing the Epoch of Reionization (PAPER)\footnote{http://eor.berkeley.edu/} \citep{parsons13} and the Murchison Widefield Array (MWA)\footnote{http://www.mwatelescope.org/} \citep{bowman13, tingay13} belong to this category. While these existing systems are limited by low sensitivity to statistical characterizations, the future low-frequency component of the Square Kilometre Array (SKA-Low)\footnote{http://www.skatelescope.org/} will have enough sensitivity to map the spatial fluctuations of the signal in tomographic images \citep{2015aska.confE..10M, 2017MNRAS.464.2234G}. The second category of systems uses a single antenna to measure the redshift evolution of the average strength of the signal. Experiments such as EDGES \citep{2010Natur.468..796B}, SARAS \citep{2015ApJ...801..138P},  BigHorns  \citep{2015PASA...32....4S}, SciHi  \citep{2014ApJ...782L...9V} and LEDA \citep{2012arXiv1201.1700G} belong to this type.

Detection of the signal in both observational setups is challenging as the signal is weaker than the astrophysical foregrounds by several orders of magnitude \citep[see e.g.,][]{2021MNRAS.500.2264H}. In addition, a long integration time is needed to reduce instrumental noise \citep[][]{ghara15c, 2017MNRAS.464.2234G, 2020MNRAS.496..739G}. Both of these challenges require excellent calibration of the radio telescope system in order to permit a detection. While significant progress towards the detection of the signal has been made, no independently confirmed detection of the 21~cm signal has been achieved thus far. However, a number of observational campaigns have provided upper limits on the strength of the signal or its fluctuations at different redshifts \cite[see e.g.,][]{paciga13, 2018ApJ...868...26C, 2019MNRAS.488.4271G, 2020MNRAS.493.1662M}. Although these upper limits are weak, the most recent ones are providing unique information about the sources at those epochs  \cite{2021MNRAS.501....1G, 2020MNRAS.498.4178M} as well as the state of the inter-galactic medium \cite{2020MNRAS.493.4728G, 2021MNRAS.503.4551G}.  

In 2018 the EDGES team claimed a detection of the redshift evolution of the global 21-cm signal \cite{EDGES2018}. The redshift-amplitude profile of the average signal as measured by the EDGES low-band instrument shows a minimum around redshift $z\approx 17$ (at a frequency of approximately 80~MHz). While such a minimum has been predicted by theoretical models of the signal, the measured absorption strength is several times stronger than the strongest signal that can be produced in a standard cosmological scenario \cite[see e.g.,][]{ghara15a, ghara15b, Pritchard07,2019MNRAS.487.1101R, 2021MNRAS.506.3717R}. While these measurements have not been independently confirmed and remain debated \cite[e.g.\ in][] {2018Natur.564E..32H, 2018ApJ...858L..10D, 2019ApJ...880...26S, 2019ApJ...874..153B}, two processes have been put forward to explain such a strong signal around redshift 17. The first one postulates that interaction between baryons and dark matter particles \cite{2018Natur.555...71B, 2018PhRvL.121a1101F, 2018Natur.557..684M, PhysRevLett.121.011102} produced a lower IGM temperature. However, limits from stellar cooling
and fifth force experiments rule out most interaction scenarios that produce excess cooling of the gas through Rutherford-like scattering with
the dark matter \cite{2018PhRvD..98j3005B, 2020MNRAS.492..634G}. The parameter space of baryon-dark matter scattering is also constrained by first star formation \cite[see e.g.,][]{2018MNRAS.480L..85H, 2019MNRAS.487.4711L}.

The other proposed process is the existence of an excess radio background (henceforth RBG) in addition to the Cosmic microwave background (CMB) around 80~MHz, the frequency corresponding to a redshift $z\approx$ 17 \cite{2018ApJ...858L..17F, 2018ApJ...868...63E, 2018PhLB..785..159F}. Such a RBG is motivated by measurements with the ARCADE \cite{2011ApJ...734....5F} and LWA1 \cite{2018ApJ...858L...9D} which have found evidence for it towards the Rayleigh-Jeans part of the CMB. The LWA1 measurement shows that the RBG at frequencies 40-80 MHz is well fitted with a power-law with spectral index -2.58$\pm$0.05 and a temperature of $603^{+102}_{-92}$ mK at 1.42 GHz.

Even though an RBG provides a viable solution for the interpretation of the EDGES low-band results, its physical origin remains uncertain. Extragalactic radio point sources are unlikely to produce a RBG strong enough to explain the measurements by EDGES. Other studies considered galactic origins such as supermassive black holes \cite{2018ApJ...868...63E} or supernovae from the first stars at redshift $z\gtrsim 17$ \cite{2019MNRAS.483.1980M}. However, the required emissivity of 1-2 GHz photons from those early sources would need to be $\sim 10^3$ times stronger than observed from local galaxies \citep{2019MNRAS.483.1980M}. An alternative to the astrophysical origins might be a RBG of cosmological origin. For example, photons produced during the decay of unstable particles into dark photons with non-zero mixing angle with electromagnetism \cite[see e.g.,][]{2018PhRvL.121c1103P} remain a viable solution for the EDGES low-band results.

Studies such as \cite{2019MNRAS.486.1763F} used the EDGES results to put constraints on a uniform RBG with a synchrotron-like spectrum that existed throughout cosmic history. However, these authors did not specify the origin of the RBG. In this paper we instead consider a physically motivated model for the RBG. We will follow the model of \cite{2018PhRvL.121c1103P} where unstable dark matter particles decay into dark photons with a small mass and non-vanishing mixing angle with electromagnetism which increase the photon count at the Rayleigh-Jeans frequencies. \cite{2018PhRvL.121c1103P} constrain the parameters of their model assuming the required radio background is a factor $\approx 2$ higher than the one which is produced by the CMB but they did not include a model for the 21-cm signal. Here, we combine a model of the 21-cm signal with the dark photon model for the RBG to constrain both the source properties at $z\approx 17$ as well as the parameters of the RBG model.

We have organised the paper in the following way. In Section \ref{sec:21cm} we describe our analytical model of the global 21-cm signal, including the model for the RBG which we use in this study. We present our results in section \ref{sec:res} and conclude in Section \ref{sec:con}. The cosmological parameters which we use in this paper are $\Omegam=0.32$, $\OmegaB=0.049$, $\Omega_\Lambda=0.68$, $h=0.67$, $\sigma_8=0.83$ and $n_{\rm s}=0.96$ \citep{Planck2015}.

\section{Methodology}
\label{sec:21cm}
\subsection{Model for the 21-cm signal}
The 21-cm signal from the \HI\ gas is measured as the differential brightness temperature against the background radiation and can be written as
\begin{equation}
 \TB   =  27 ~x_{\rm HI} (1+\delta_{\rm B}) \left(\frac{\OmegaB h^2}{0.023}\right) \sqrt[]{\frac{0.15}{\Omegam h^2}\frac{1+z}{10}}  \left(1-\frac{\TRAD}{\TS} \right)\,\rm{mK},
\label{eq_tb}
\end{equation}
where $x_{\rm HI}$, $\delta_{\rm B}$ and $\TS$ denote the neutral fraction, density contrast, the spin temperature of the hydrogen gas respectively. $\TRAD$ is the temperature of the radio background.  $\TRAD$ becomes  the CMB temperature $\TCMB=2.725 \times (1+z)$ K  in absence of a RBG. Note that the spin-temperature calculation in our model depends on collisional coupling, $\lya$ coupling and the scattering with the CMB photons. All these coupling strengths are dependent on $\TRAD$ (see e.g. \cite{2019MNRAS.486.1763F}). 

We adopt an analytic approach as described below to model the expected 21-cm signal in the presence of spin temperature fluctuations. The analytical model \citep[previously used in ][]{2019JCAP...04..051N, 2020MNRAS.492..634G} derives from previous works such as \citep{Pritchard07, 2005ApJ...630..643M}. It incorporates $\lya$, UV, and X-ray photons from sources of radiation which are presumed to have formed from gas associated with dark matter halos. The number density of dark matter halos at a given redshift is determined using the Press-Schechter halo mass function. We assume that star formation only occurs inside dark matter halos with virial temperatures above $T_{\rm vir}$. We vary $T_{\rm vir}$ in this study.

The model estimates the volume averaged ionization fractions of the highly ionized \HII ~regions ($x_i$) and largely neutral gas in the IGM outside these \HII ~regions ($x_e$). We assume the temperature of the ionized \HII ~regions to be $\sim 10^4$~K but estimate the gas temperature ($\TK$) of the largely neutral medium outside the \HII ~regions from the relevant the heating and cooling processes. 

The average number density of UV, X-ray, and $\lya$ photons in the model follows the mass fraction of collapsed objects, $f_{\rm coll}$. The rate of emission of UV photons per baryon is given by
\begin{equation}
\Lambda_i = \zeta \frac{ {\rm d} f_{\rm coll}}{ {\rm d} t}.
\end{equation}
The ionization efficiency parameter $\zeta = N_{\rm ion}\times f_{\rm esc}\times f_{\star}$ depends on the average number of ionizing photons per baryon produced in the stars ($N_{\rm ion}$), the star formation efficiency ($f_\star$) and the escape fraction of the UV photons ($f_{\rm esc}$). All these quantities are uncertain during the Cosmic Dawn. Their effects on the ionization state of the IGM are also degenerate. We therefore fix $N_{\rm ion}=4000$, which corresponds to population II types of stars, and $f_{\rm esc}=0.1$, and only vary $f_\star$ for modelling reionization. We note that ionization levels remain very low for the redshifts that we consider here and thus the ionization state of the IGM does not have a significant impact on the global 21-cm signal. 

Our model also includes X-ray sources which can partially ionize and heat the neutral IGM at large distances from the \HII\ regions which form around the sources. We follow \citep{Pritchard07} to include their impact. This approach assumes that the emissivity of X-ray photons from the sources follows the star formation rate density. The comoving photon emissivity for X-ray sources is modelled as
\begin{equation}
\epsilon_X(\nu, z) = \epsilon_X(\nu) \left(\frac{\rm SFRD}{\MSUN~ \rm yr^{-1} ~\rm Mpc^{-3}} \right),
\end{equation}
where SFRD is the star formation rate density. We model the SFRD as
\begin{equation}
{\rm SFRD} (z) = 
\bar{\rho}_b^0 ~f_\star ~\frac{d}{dt}f_{\rm coll}(z),
\end{equation}
where $\bar{\rho}_b^0$ is the cosmic mean baryon density at $z=0$.
We use an X-ray spectral distribution given by 
\begin{equation}
\epsilon_X(\nu) \propto \frac{L_0}{h\nu_0}\left(\frac{\nu}{\nu_0}\right)^{-\alpha_X-1}
\label{eq:epX}
\end{equation}
with $L_0=f_X\times 10^{41} ~\rm erg ~s^{-1} ~Mpc^{-3}$, $h\nu_0 = 1 ~\rm keV$. We choose the proportional constant in equation \ref{eq:epX} such that the rate of emission of energy in X-ray band (0.1-10 keV in our case) per unit volume per unit SFRD is $L_0$. It is therefore equivalent to assuming that stellar populations produce a total of $f_X\times 3.156\times 10^{48}$ erg $\MSUN^{-1}$ in the X-ray band. We fix the X-ray spectral index $\alpha_X$ to 0.5 but vary the X-ray parameter $f_X$.  

The $\lya$ background has a crucial role in determining the 21-cm signal through coupling the spin-temperature of atomic hydrogen with the gas temperature through Wouthuysen-Field coupling \cite{hirata2006lya}.
Here we used the methodology of \citep{2006MNRAS.372.1093F} to estimate the average $\lya$ photon flux. We assume a power-law spectrum  $\epsilon_s(\nu) = f_{\alpha} A_{\alpha} \nu^{-\alpha_{s}-1}$ between $\lya$ and $\lyb$ and between $\lyb$ and the Lyman limit, where the power-law indices $\alpha_{s}$ can differ between these two spectral regimes. The spectral index $\alpha_s$ between $\lya$ and $\lyb$ is taken to be 0.14 which corresponds to population II type sources. The normalization factor $A_{\alpha}$ is estimated such that the number of $\lya$ photons per baryon in the range $\lya$--$\lyb$ is 6520 for $f_{\alpha}=1$. The spectral index in the range $\lya$--Lyman limit is adjusted so that the total number of photons per baryon for this wavelength regime is 9690.

Our model for the 21-cm signal thus has three free parameters, namely the star formation efficiency $f_\star$, the X-ray efficiency $f_X$ and minimum virial temperature for star forming halos, $T_{\rm vir}$. These control the $\lya$ coupling, X-ray heating and the number density of these sources, respectively.


\subsection{Model for the excess radio background}
\label{sec:RBG}
We model the RBG following \citep{2018PhRvL.121c1103P}.
The first part of this model consists of the decay of long-lived unstable scalar particles $a$ into light vector particles $A^{'}$, which are often called dark photons. If we assume the life time of the scalar particles $\tau_{a} \gg \tau_{U}$ where $\tau_{U} \approx 13.8\times 10^{9}$ yr is the age of the Universe, then dark matter is a natural candidate for this scalar field. The energy spectrum of $A^{'}$ at redshift $z$ due to the decay $a \rightarrow 2A^{'}$ can be written as \citep{2018PhRvL.121c1103P},
\begin{equation}
    \frac{dn_{A^{'}}}{dw}(w, z) =  \frac{2\Omega_a \rho_c (1+z)^3}{m_a \tau_a w H(\alpha-1)} \Theta(\alpha-1-z)
    \label{eq:deacy}
\end{equation}
where $\Omega_{a} = \Omegam$ in our case, $\rho_{c}$ is the critical density, $H$ is Hubble rate evaluated at redshift $\alpha-1$ where $\alpha = m_{a}(1+z)/(2w)$.  Here, $m_a$ is the mass of the scalar particles and $w$ is the photon energy.

The second part of this model is the mixing of the dark photons $A^{'}$ with ordinary photons $A$. Cosmological $A^{'}\leftrightarrow A$ oscillation is the most significant under resonance conditions, $m_{A^{'}}=m_A(z) \simeq 1.7 \times 10^{-14} \times (1+z)^{3/2} X^{1/2}_e(z)$ eV where $m_A^{'} \rightarrow 0$  is the mass of the dark photons and $m_A(z)$ is the plasma mass of the photons at redshift $z$ \citep[][]{Mirizzi_2009, Kunze_2015}. The redshift evolution of the average ionization fraction $X(z)$ is derived from {\sc recfast} \cite{2000ApJS..128..407S}. The probability of oscillation at the resonance can be written as \citep[][]{Mirizzi_2009}
\begin{equation}
    P_{A\rightarrow A^{'}} = P_{A^{'}\rightarrow A} \simeq 1- \exp{\left( -\frac{\pi \epsilon^2 m^2_{A^{'}}}{w} \times \left|\frac{d {\rm log}~ m^2_A}{dt}\right|^{-1}_{t=t_{\rm res}}\right)}
\end{equation}
where $\epsilon$ is the mixing angle of the dark and ordinary photon oscillation, $w$ is the energy of the photon, $t_{\rm res}$ is the time of resonance for $m_{A^{'}}$. This modifies the number count of CMB photons as \citep{2018PhRvL.121c1103P},

\begin{equation}
    \frac{dn_A}{dw} \rightarrow \frac{dn_A}{dw} \times (1-P_{A\rightarrow A^{'}}) + \frac{dn_A^{'}}{dw} \times P_{A^{'}\rightarrow A}.
\end{equation}

Our physical model for the RBG of cosmological origin thus has four free parameters, namely $m_a$, $\tau_a$, $m_{A^{'}}$ and $\epsilon$. The decay of dark matter can produce a significant impact on the Rayleigh-Jeans tail of the CMB if $m_a$ remains between $10^{-5}$ to $10^{-1}$ eV. For $m_a \lesssim 10^{-5}$ eV, the produced photons due to the decay of dark matter are too soft to have an impact on the 21-cm signal at $z\approx 17$. Assuming that all cold dark matter particles decay into dark photons, the lifetime $\tau_a$ could be infinitely large. We choose $\tau_a$ range between 12 to $10^7$ $\tau_U$. The larger values will have insignificant effects on the RJ tail of the CMB while the lower bound is taken from \cite{2016JCAP...08..036P} which only relies on the CMB to obtain the constraint. We choose $m_{A^{'}}$ range $10^{-14}-10^{-9}$ eV for which the time when the resonance occurs falls between recombination and redshift of our interest. The bounds on $\epsilon$ are sensitive to the bounds on $m_{A^{'}}$. We vary $\epsilon$ in the range $10^{-10}-10^{-4}$ which is the most relevant range for this study.

\subsection{Bayesian Framework}
\label{sec:frame}
The main aim of this work is to infer astrophysical and particle physics information using the global 21-cm signal vs redshift profile as extracted from EDGES low-band data. As mentioned in the earlier sections, we have a total of seven free parameters in our combined model, three related to astrophysical radiation sources and four related to the RBG model. Given a set of these seven parameters ($\theta$), the above mentioned method generates a $\TB$ vs $z$ profile which we later compare to the measured EDGES profile in a Bayesian framework. We couple the above mention code to the {\sc cosmomc} Markov chain Monte Carlo (MCMC) generic subroutine \citep{2002PhRvD..66j3511L} to explore the seven-dimensional parameter space. Note that we do not consider all the frequency channels of EDGES observation, instead, we consider redshifts 21.2, 20., 19.4, 18.9, 17.3, 16.1, 15.2, 14.9, and 14.4 to estimate the $\chi^2(\theta) = \sum_{i} \left(\frac{\delta T_{\rm b,m}(z_i, \theta)-\delta T_{\rm b,o}(z_i)}{\sigma_{\rm o}}\right)^2$ where $\delta T_{\rm b,m}$ and $\delta T_{\rm b,o}$ are the modelled $\TB$ and the observed $\TB$ respectively.  We assume $\sigma_{\rm o}$ is 25 mK which is consistent with \citep{2018Natur.564E..32H}. We use $-\chi^2(\theta)$ as the likelihood in the MCMC algorithm.

The parameter ranges explored in this study are given in Table \ref{tab1}. We use flat priors over the ranges of the parameter space. We use one additional flat prior on $\epsilon$-$m_{A^{'}}$ which is based on the findings of COBE-FIRAS \cite{1996ApJ...473..576F}. The observed data from COBE-FIRAS has provided strong constraints on the values of $\epsilon$ as a function of $m_{A^{'}}$ which exclude a significant part of the parameter space with high values of $\epsilon$ \cite{2009JCAP...03..026M}. 

\begin{table}
\centering
\small
\tabcolsep 6pt
\renewcommand\arraystretch{1.5}
   \begin{tabular}{c c c c c c c c}
\hline
\hline
Scenarios & log($f_\star$) & log($f_X$) & log($\frac{T_{\rm vir}}{K}$)  & log($\frac{m_a}{\rm eV}$) & log($\frac{\tau_a}{\tau_U}$) & log($\frac{m_{A^{'}}}{\rm eV}$) & log($\epsilon$)	 \\

\hline
\hline
$S_1$ & -1.   & 0.0   &  4.0     & -3.0	  & 2.0	& -11.5 & -6.4		\\
$S_2$ & \textit{-0.3}   & 0.0   &  4.0     & -3.0	  & 2.0	& -11.5 & -6.4		\\
$S_3$ & -1.   & \textit{1.0}   &  4.0     & -3.0	  & 2.0	& -11.5 & -6.4		\\
$S_4$ & -1.   & 0.0   &  \textit{4.5}     & -3.0	  & 2.0	& -11.5 & -6.4		\\
$S_5$ & -1.   & 0.0   &  4.0     & \textit{-2.5}	  & 2.0	& -11.5 & -6.4		\\
$S_6$ & -1.   & 0.0   &  4.0     & -3.0	  & \textit{3.0}	& -11.5 & -6.4		\\
$S_7$ & -1.   & 0.0   &  4.0     & -3.0	  & 2.0	& \textit{-12.0} & -6.4		\\
$S_8$ & -1.   & 0.0   &  4.0     & -3.0	  & 2.0	& -11.5 & \textit{-7.5}		\\
\hline
MCMC & [-4, -0.3]   & [-4, 4] & [3, 5]   &  [-5, -1]     & [1.08, 7]	  & [-14, -9]	& [-10, -4]		\\
\hline
\hline
\end{tabular}
\caption[]{Sets of seven parameters of the eight different scenarios considered in Section \ref{sec:scenario}. The bottom row shows the range of those seven parameters as used in the MCMC analysis in section \ref{sec:mcmc}.}
\label{tab1}
\end{table}

\begin{figure}[t]
 \centering
\includegraphics[width=\textwidth]{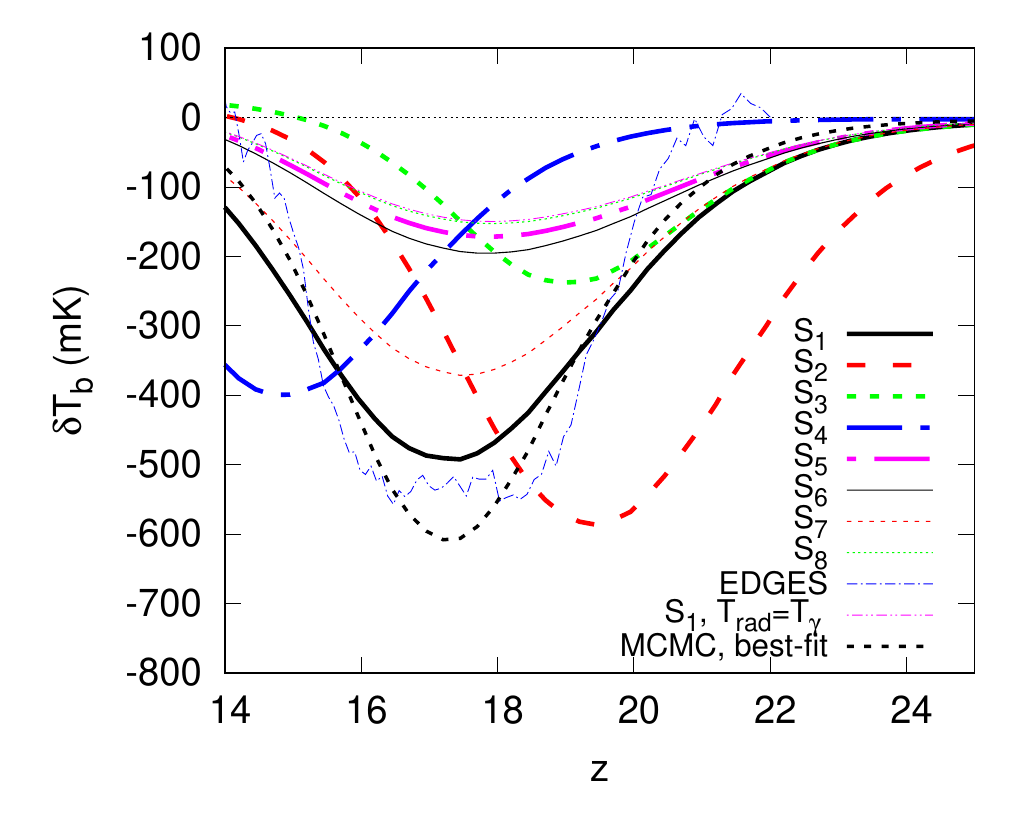}
    \caption{Average of the differential brightness temperature as a function of redshift. These different models are considered in section \ref{sec:scenario}. The thin dot-dashed curve shows the measured $\TB$ vs redshift profile by EDGES-low observation. The thin double-dot dashed curve shows the $\TB$ profile when the astrophysical parameters are the same as $S_1$ and $\TRAD=\TCMB$. The black dotted curve shows the $\TB$ profile for the best-fit parameter values from the MCMC analysis in section \ref{sec:mcmc}.}
   \label{image_RBG_param}
\end{figure}

\begin{figure}[t]
 \centering
\includegraphics[width=\textwidth]{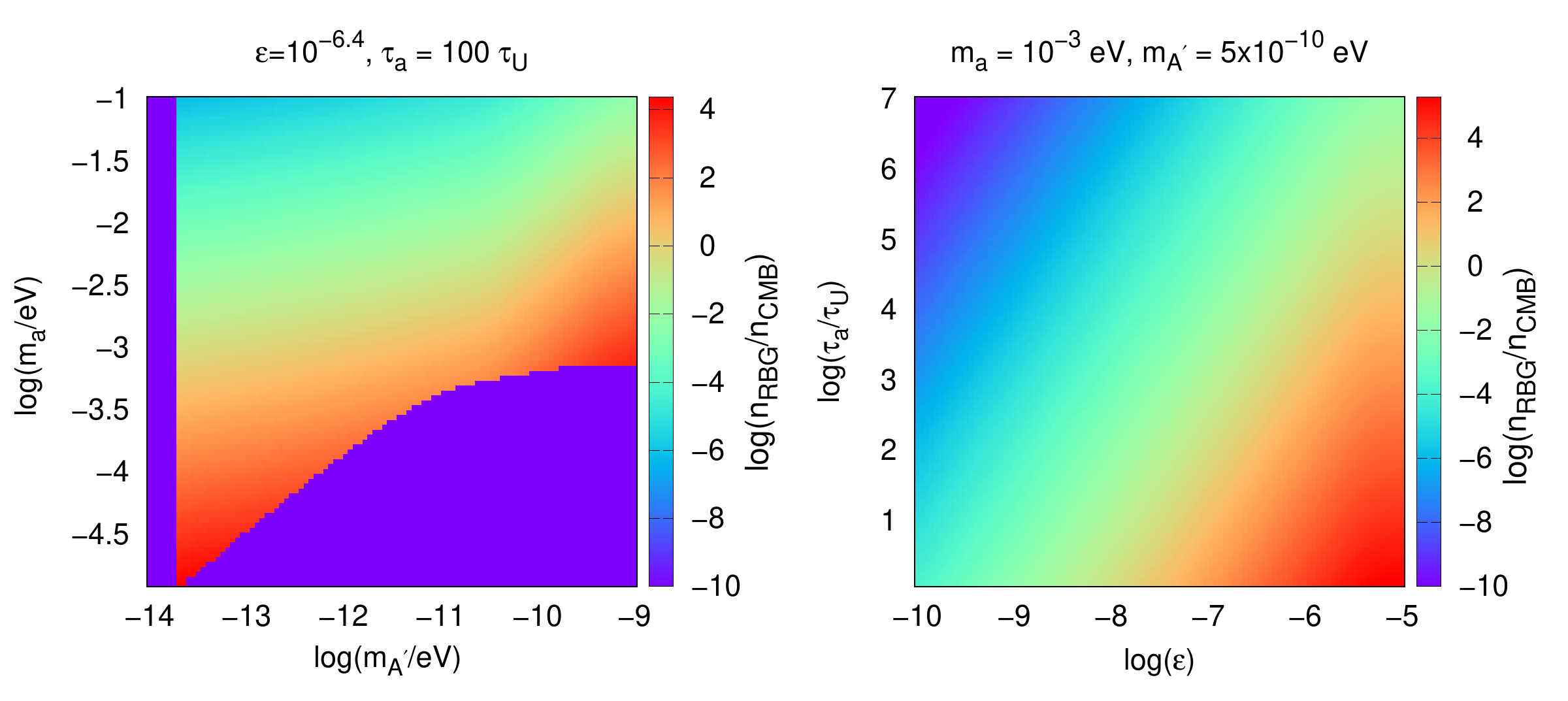}
    \caption{The color bar shows the logarithm of the ratio of the contribution from the RBG model and the CMB at a wavelength of 21-cm at redshift 17. Two panels show this as a function of parameters of the RBG model.}
   \label{image_RBG}
\end{figure}

\section{Results}
\label{sec:res}

\subsection{Exploratory scenarios}
\label{sec:scenario}
We first explore a few scenarios/models to understand the impact of each of the parameters on the global 21-cm signal from the Cosmic Dawn. In total we consider eight scenarios with different combinations of the seven parameters. The parameter values of those scenarios are listed in Table  \ref{tab1}. The redshift evolution of the average differential brightness temperature for each of these scenarios is shown in Figure \ref{image_RBG_param}.

Our fiducial model $S_1$ corresponds to $f_\star=0.1, f_X=1, T_{\rm vir}=10^4 ~{\rm K}, m_a = 10^{-3} ~{\rm eV}, \tau_a = 100\times \tau_U, m_{A^{'}}=10^{-11.5} ~{\rm eV}, \epsilon= 10^{-6.4} $. For these parameter values, the corresponding $\TB$ vs redshift profile has an absorption trough around redshift $\approx 17.5$ with an amplitude $\approx 0.5$ K. For reference Figure \ref{image_RBG_param} also shows the $\TB(z)$ profile for the same astrophysical parameters as in scenario $S_1$ but without a RBG (see thin double-dot dashed curve). In that case the minimum $\TB$ remains $\approx -150$ mK at a redshift $\approx 18$. One can also see that the $\TB$ profile of model $S_1$ does not quite match with the measured $\TB$ profile from EDGES (see dot-dashed curve). Especially, the width of the $\TB$ profile of the $S_1$ model is larger than the width of the measured $\TB$ profile by EDGES-low. One might guess that a delayed star formation history and more efficient heating of the gas in the IGM is required to obtain a $\TB$ profile that provides a better fit to the EDGES measurements than the $S_1$ model does.

Scenario $S_2$ only differs from $S_1$ in its value for $f_\star$. Its larger $f_\star$ value of $\approx 0.5$ increases the $\lya$ coupling as well as the heating due to the X-rays. This results in a shift of the trough to higher redshift compared to the $S_1$ scenario. Also, the depth of the trough is enhanced.

We choose a higher value of $f_X=10$ for scenario $S_3$ while all other parameters have the same value as in $S_1$. This causes more efficient X-ray heating from the beginning of Cosmic Dawn which shifts the trough to higher redshift while reducing its depth.

Scenario $S_4$ has a larger $T_{\rm vir}$ at $10^{4.5}$ K as compared to $S_1$ model. Removing the lower mass halos results in a delay in the global star formation  and an overall reduction of the effects of stars. This shifts the abosrption trough to lower redshifts and decreases its depth.

After having explored the impact of the astrophysical parameters, we now turn to the parameters of the RBG model. Scenario $S_5$ is identical to $S_1$ except for a higher $m_a$ value of $10^{-2.5}$ eV. This results in a smaller number density of the dark photons (see Equ \ref{eq:deacy}). This reduces the depth of the absorption trough and shifts it to a higher redshift.

Increasing the value of $\tau_a$ also causes a decrease of number density of the dark photons and thus has a similar effect as increasing $m_a$, as can be seen in the $\TB$ profile for scenario $S_6$ which has $\tau_a=10^3 ~\tau_U$, a value 10 times larger than used in $S_1$.

Scenario $S_7$ has a smaller $m_{A^{'}}$ value of $10^{-12}$ eV compared to $m_{A^{'}}=10^{-11.5}$ eV in $S_1$. This causes a smaller $P_{A\rightarrow A^{'}}$ value as $P_{A\rightarrow A^{'}} \propto m^2_{A^{'}}$. A similar effect can be achieved by reducing the value of $\epsilon$ as is done in scenario $S_8$ where $\epsilon$ is a bit more than 10 times lower than in $S_1$. This leads to shallower absorption troughs which are shifted to larger redshifts. The reduction in the RBG for scenario $S_8$ is such that its profile resembles the one in which the RBG is entirely absent.

Figure \ref{image_RBG} illustrates the impact of the four RBG parameters on the RBG. The color bar shows the logarithm of the ratio of the contribution of the RBG  and the CMB at a wavelength of 21-cm at redshift 17. One can see that the ratio increases for smaller $m_a$ and larger $m_A^{'}$. The region in purple in the right bottom corner of the left panel of Fig \ref{image_RBG} represents a part of the parameter space for which the produced photons are too soft and do not have an impact on the 21-cm signal at $z\approx 17$. For $m_{a} \lesssim 10^{-13.7}$ eV (purple region on the left hand side of the same panel), the redshift of resonance is smaller than $z\approx 17$ which makes $P_{A^{'}\rightarrow A} = 0$. Note that we fix $\epsilon=10^{-6.4}$ and $\tau_a=10^2 ~\tau_U$ for the left panel.

We vary $\epsilon$ and $\tau_a$ in the right panel of Fig \ref{image_RBG} while we fix $m_a=10^{-3}$ eV and $m_{A^{'}}=5\times 10^{-10}$ eV. The ratio of the contribution from the RBG and the CMB increases for smaller $\tau_a$ and $\epsilon$ values. These panels give some idea about which parts of the 4D parameter space of the RBG cannot satisfy the measured $\TB$ profile. However, it is difficult to pinpoint values of the ratio of the contribution from the RBG and the CMB that are required for the measured $\TB$ profile as that will crucially depend on the three astrophysical parameters. Thus, we will explore the full seven-dimensional parameter space in the next section to find the part of the parameter space that best agrees with the observation.  

\begin{table}
\centering
\small
\tabcolsep 6pt
\renewcommand\arraystretch{1.5}
   \begin{tabular}{c c c c c c c c c}
\hline
\hline
Parameters & Explored range & Mean & Standard Deviation & Best fit  & 68$\%$ limits & 95$\%$ limits	 \\

\hline
\hline
$\mathrm{log}(f_\star)$ & [-4, -0.3]   & -0.38   &  0.08  & -0.3    & [-0.4, -0.3]	& [-0.55, -0.3]		\\
$ \mathrm{log}(f_X)$ & [-4, 4]   & 0.26   &  0.14   & 0.14  	  & [0.12, 0.34]	& [0.01, 0.54]		\\
$ \mathrm{log}(\frac{T_{\mathrm{vir}}}{K})$ & [3, 5]   &  4.41  &  0.04  & 4.37   	  & [4.34, 4.4]	& [4.28, 4.43]		\\
$\mathrm{log}(\frac{m_a}{\mathrm{eV}})$ & [-5, -1]   & -3.03  &  0.53  & -3.84  	  & [-3.5, -2.4]	& [-4.1, -2.1]		\\
$\mathrm{log}(\frac{\tau_a}{\tau_U})$ & [0.5, 7]   & 2.27   &  0.9  & 1.48     & [1.1, 2.7]	& [1.07, 4.0]		\\
$\mathrm{log}(\frac{m_{A^{'}}}{\mathrm{eV}})$ & [-14, -9]   & -11.3   &  1.1 & -12.76	  & [-12.2, -10.0]	& [-13.2, -9.4]		\\
$\mathrm{log}(\epsilon)$ & [-10, -4]   & -6.3   &  0.64  & -7.38	  & [-7., -5.6]	& [-7.5, -5.2]		\\
\hline
\hline
\end{tabular}
\caption[]{68 and 95 percent credible intervals on the seven parameters used to model the $\TB$ in this study. These limits are obtained from the measurements of $\TB$ vs redshift profile by EDGES-low band observation. }
\label{tab2}
\end{table}

\begin{figure}[t]
 \centering
\includegraphics[width=\textwidth]{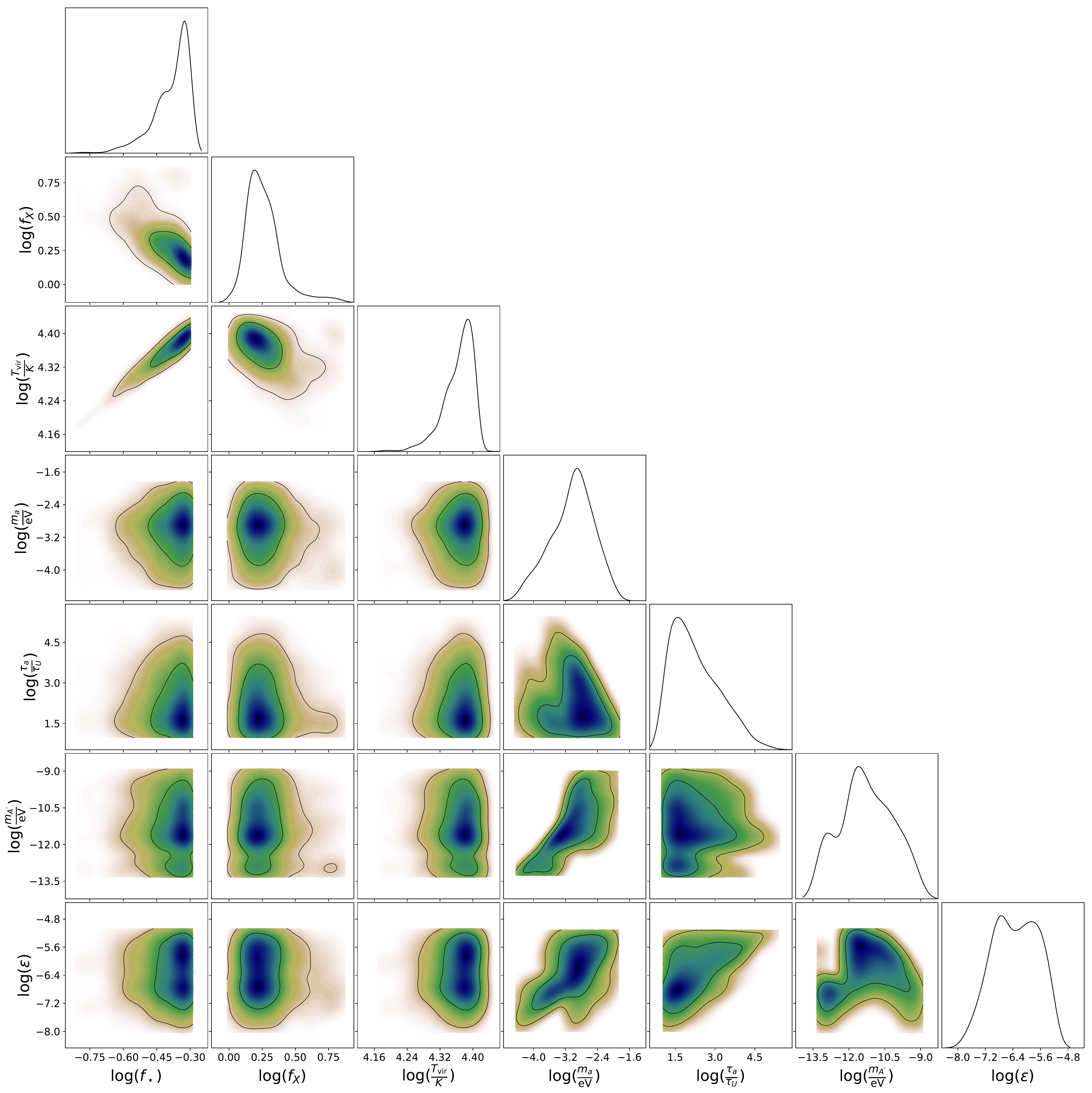}
    \caption{Constraints on the parameters of this study from the MCMC analysis using the EDGES observation. The color-bar shows the probability that models are valid. The solid and dashed curves show the 68 and 95 percent credible intervals of the
consistent models. The diagonal panels show the 1D marginalized probability distribution of the parameters.}
   \label{image_mcmc}
\end{figure}


\subsection{MCMC results}
\label{sec:mcmc}
The details of the framework to explore the seven-dimensional parameter space are given in section \ref{sec:frame} and the range of the parameters explored in the MCMC analysis is listed in Table \ref{tab1}. Additional priors to this MCMC analysis are the allowed parameter space of $\epsilon$ and $m_{A^{'}}$ derived from the COBE-FIRAS results and the age of the decaying dark matter particles $\tau_a$, i.e. $\tau_a>12 \tau_U$, as obtained in \cite{2016JCAP...08..036P}. 

Figure \ref{image_mcmc} shows the outputs of the MCMC analysis and the 68 and 95 percent credible intervals on the parameters are listed in Table \ref{tab2}. The data is found to prefer a high star formation efficiency, reaching the maximum value of our prior range. The 68 percent credible interval for $f_X$ is [1.3, 2.2], while that for $T_{\rm vir}$ is $[2.2, 2.5] \times 10^4$ K. The limit on $T_{\rm vir}$ indicates a scenario where the star formation occurs inside the massive halos only which suggests a delayed star formation history. This is expected as a smaller value of $T_{\rm vir}$ shifts the $\TB$ vs redshift profile towards a higher redshift (as we have shown by comparing scenario $S_1$ and $S_4$ in section \ref{sec:scenario}). However, a larger value of $T_{\rm vir}$ corresponds to a smaller value of $f_{\rm coll}$ which tends to produce a weaker signal due to a weaker $\lya$ coupling. In the best-fit star formation scenario, this is compensated for by a larger value of $f_\star$ that enhance the strength of the signal (see the comparison of scenarios $S_1$ and $S_2$ in section \ref{sec:scenario}). On the other hand, the limits of $f_X$ indicate efficient X-ray heating which is required for the quick transition of the mean signal from absorption to emission at redshift $\lesssim 17$ (see the comparison of scenarios $S_1$ and $S_3$ in section \ref{sec:scenario}).

It is straightforward to understand that the limits on the astrophysical parameters as obtained from the MCMC analysis are mainly determined by the shape of the $\TB$ vs redshift profile. As we find that the best-fit scenario indicates a high value for the $f_X$ parameter, one might expect a non-negligible impact of X-ray heating on the gas at the redshifts of our interest. On the other hand, the $\lya$ coupling is expected to be not very strong as the MCMC analysis indicates a higher value of $T_{\rm vir}$ which means a smaller $f_{\rm coll}$ value. These two effects result in a weaker 21-cm signal compared to the scenario in which the IGM is completely unheated in the presence of a strong $\lya$ background. Thus, one might expect a stronger RBG will be required as compared to \cite{2018PhRvL.121c1103P} to compensate for the stronger X-ray heating and weaker $\lya$ coupling at these redshifts.

The 68 percent credible intervals for the four parameters of the RBG, namely $m_a$, $\tau_a$, $m_{A^{'}}$ and $\epsilon$, are $[10^{-3.5}, 10^{-2.4}]$ eV, $[10^{1.1}, 10^{2.7}]\times \tau_U$, $[10^{-12.2}, 10^{-10.0}]$ eV and $[10^{-7}, 10^{-5.6}]$. The effective temperature of the radio background  corresponding to the best fit parameter values at $z\approx 17$ is $\TRAD\approx 5.7\times \TCMB$. This is consistent with the $\TRAD(z\approx 17)\gtrsim 2.9\times \TCMB$ as found in \cite{2019MNRAS.486.1763F}. This also shows that the data prefer a much stronger RBG than the one assumed in \cite{2018PhRvL.121c1103P}, namely $\TRAD\approx 2\times \TCMB$.

We show the $\TB$ vs redshift profile for the best-fit values of the parameters in Fig.~\ref{image_RBG_param} (thick black dotted curve). It is obvious that it is still difficult to reproduce the exact shape of the measured curve with the source model used in this study. To obtain a flat-bottomed $\TB$ profile will need exotic source models such as the one discussed in \cite{2020MNRAS.496.1445C}. It is worth pointing out that even though the source models differ, our results are consistent with $\TRAD(z\approx 17)\approx 5.3\times \TCMB$ as obtained in the best-fit model of \cite{2020MNRAS.496.1445C}.    

 \section{Discussion \& conclusions}
\label{sec:con}

In this work we interpret the claimed EDGES detection of a deep 21-cm absorption signal around $z\approx 17$ in terms of RBG produced by dark matter particles decaying into dark photons that later oscillate into ordinary photons due to a nonvanishing mixing angle. This physical model can enhance the photon number density in the Rayleigh-Jeans tail of the CMB that remains unconstrained till now. Such an enhanced radio background might be an explanation for the EDGES-low band results which show an absorption signal a few factors stronger than theoretical predictions based on standard physics.

The main aim of this work is to find constraints both on the properties of the astrophysical sources of radiation around redshift $\approx 17$ as well as the particle physics parameters of the dark photon model for the RBG. For this, we explored a seven-dimensionial parameter space in which three parameters describe the astrophysical sources, namely the star formation efficiency $(f_\star)$, X-ray heating efficiency $f_X$ and the minimum virial temperature $T_{\rm vir}$ of dark matter halos in which star formation occurs. The other four parameters determine the RBG, namely the mass of the decaying dark matter particles $(m_a)$, the lifetime of the dark matter particles $(\tau_a)$, the mass of the dark photons $(m_{A^{'}})$ and the mixing angle between dark and ordinary photon oscillation ($\epsilon$).

We explored this seven-dimensional parameter space using a Bayesian framework to obtain constraints on the parameters using the $\TB$ vs redshift profile as measured by the EDGES-low band experiment. Our main findings are as follows (numerical values correspond to 68 percent credible intervals):
\begin{itemize}
    \item We find that a higher star formation efficiency (0.5 in this case)  and an intermediate value of $f_X\approx 1-2$ which means emission of X-ray energy per stellar mass of $4-7 \times 10^{48} ~\rm erg$ $\MSUN^{-1}$ is required to to produce efficient heating at $z\lesssim 17$. At the same time, one needs a star formation suppression at dark matter halos with virial temperature $\lesssim 2\times 10^4$ K to produce a late start of the $\TB$ profile.

    \item In addition, one needs a strong radio background which provides tighter constraints on the parameters of the radio background model. The 68 percent credible intervals on $m_a$, $\tau_a$, $m_{A^{'}}$ and $\epsilon$ are $[10^{-3.5}, 10^{-2.4}]$ eV, $[10^{1.1}, 10^{2.7}]\times \tau_U$, $[10^{-12.2}, 10^{-10}]$ eV and $[10^{-7}, 10^{-5.6}]$ respectively. 
    
    \item The best fit parameter values suggest a radio background with temperature $\approx 5.7\times \TCMB$.  
\end{itemize}

The derived constraints on the astrophysical source parameters are in disagreement with those from studies which assume that the radio background only consists of the black body radiation of the CMB and instead use an an excess cooling mechanism to explain the EDGES results. Those scenarios prefer a lower star formation efficiency combined with a higher cooling rate.

Our results show that the EDGES-low band observation alone excludes a large part of the parameter space for a RBG due to the decay of dark matter into dark photons that later oscillate into ordinary photons. However, a significant part of the parameter space still remains valid. Future probes such as PIXIE \cite{2011JCAP...07..025K} or PRISM \cite{2014JCAP...02..006A} will provide stronger constraints on the parameters of the radio background as well as on the astrophysical parameters. 

The findings of this study are based on the assumption that the dark matter halo mass function (HMF) follows a Press-Schechter (PS) \cite{Press_1974} form. However, as for example shown in \cite{2013MNRAS.433.1230W}, the PS-HMF overestimates the number of low mass halos and underestimates the number of rare high mass halos compared to results from N-body simulations. At the redshifts we are considering here, halos with $T_\mathrm{vir}\gtrsim 2\times 10^4$ K are rare and their abundance is therefore likely underestimated by the PS-HMF. The HMF from the ellipsoidal collapse model from Sheth \& Tormen (ST) \cite{Sheth1999} has been shown to overestimate the number of rare high mass halos compared to numerical results. We therefore repeated our analysis using the ST-HMF to investigate the impact of the choice of HMF. 
For the ST-HMF the 68 percent credible interval limits for log($f_\star$), log($f_X$), log($\frac{T_{\rm vir}}{\rm K}$), log($\frac{m_a}{\rm eV}$), log($\frac{\tau_a}{\tau_U}$), log($\frac{m_{A^{'}}}{\rm eV}$) and log($\epsilon$) change to [-0.6, -0.46], [0.45, 0.78], [4.7, 4.9], [-4.5, -2.1], [1.1, 5.3], [-13.2, -9] and [-8.1, -5.1] respectively. The best fit values of these parameters are -0.47, 0.73, 4.77, -3.9, 1.59, -12.96 and -7.17 respectively. Clearly the choice of HMF has a considerable impact on the results, which is not surprising if one realises that $f_{\rm coll}(T_{\rm vir}=10^4 \rm K, z=17)$ for the ST HMF is approximately a factor 4 higher than for the PS HMF. With a similar analogy, the limits on our model parameters are expected to be in between the limits from the PS and ST HMF cases if the HMF follows the form as given in \cite{2013MNRAS.433.1230W}.

Besides HMF, the obtained constraints on the parameters of this study are also limited by the uncertainties on the type of sources present at those redshifts. The source model of this study assumes only population II type stars. While it is expected that these stars dominate the contribution to the overall photon emissions at the redshifts of our interest, the contribution from population III type stars might be also important \cite{2019MNRAS.486.1763F, 2019ApJ...877L...5S, 2020MNRAS.495..123Q}. The emission of $\lya$, X-ray and radio photons from population III stars and their remnants can have a significant impact on the shape of the brightness temperature profile at these redshifts \cite[see, e.g.,][]{2020MNRAS.493.1217M} and thus, will require a smaller contribution to the overall radio ($\lya$ and X-ray) background from the RBG (population II star) model. 

One needs to keep in mind that the results of this study are based on the simple-minded source model which assumes the star formation efficiency ($f_\star$) is constant over the range of dark matter halo masses ($M_{\rm halo}$) and redshifts relevant for the EDGES signal. In general, $f_\star$ varies with hosting dark matter halo mass and redshift. Motivated by the observations of the high-redshift ($z\lesssim 8$) faint galaxy population \citep[e.g.,][]{2015ApJ...799...32B}, the halo mass dependencies on $f_\star$ are often characterised by a power-law such as $f_\star \propto M^{\alpha_\star}_{\rm halo}$ with positive value of $\alpha_\star$. This suppresses (enhances) star-formation in low-mass (high-mass) halos. The mass-dependent $f_\star$ also allows molecular-cooling low-mass halos to form Population III stars at low efficiency which is consistent with theoretical predictions \citep[see e.g.,][]{2007MNRAS.382..945T, 2019ApJ...871..206S}. Under such a sophisticated model for star formation as used in \cite{2020MNRAS.495..123Q}\footnote{Note that the constraints in \cite{2020MNRAS.495..123Q} are based on the position of the 21-cm absorption trough at 78 MHz frequency, while we consider the entire shape of the brightness temperature profile within the redshift range of EDGES-low.} the 68 percent credible limits on $\Tvir$ are lower than what is obtained in this study. Depending on different feedback mechanisms such as Lyman-Werner feedback, radiative feedback, etc. the star formation efficiency can also evolve with redshift, especially when the population III star contribution is taken into account \citep[see, e.g.,][]{2020ApJ...897...95V, 2018MNRAS.479.4544M, 2020MNRAS.497.2839L}. The limits on our source parameters are expected to change in case $f_\star$ changes rapidly within the redshift window considered in this study (as for example considered in \cite{2020MNRAS.496.1445C}). For a rapidly evolving $f_\star$, a scenario with smaller values of $\Tvir$ and $f_\star$ (similar to the one inferred in \cite{2020MNRAS.495..123Q}) can also produce a $\TB$ profile that matches the observed one but the shape of the modelled $\TB$ profile will be mainly determined by how $f_\star$ evolves with redshift.

Our limits on the RBG are also limited by the uncertainties in the HMF and the star formation scenarios. For example, the best fit parameter values for the ST-HMF and constant $f_\star$ suggest a radio background with temperature $\approx 10.6\times \TCMB$ at $z\sim 17$ which is $\approx 2$ factor larger than what we obtain for the PS-HMF. The required RBG will also differ with the X-ray heating scenario. For example, in a scenario such as the one used in \cite{2022MNRAS.509..945D} where X-ray heating only is active for $z\lesssim 17$ and an evolving SFR saturates the $\lya$ coupling at $z\approx 17$, the required RBG will be the same as reported in \citep{2018PhRvL.121c1103P}.

Lastly, the constraints obtained in this study are strongly dependent on the model for dark photons. They are of course also dependent on the priors, especially on the priors for $m_{A^{'}}$ and $\epsilon$ space \cite{2009JCAP...03..026M} which are based on $A\rightarrow A^{'}$ transitions and thus in a sense are the least model-dependent priors. Alternative models for non-relativistic dark photons such as dark photon dark matter models \cite{2020PhRvD.101f3030M, 2020JHEP...06..132W} in which dark photons comprise dark matter, estimate stronger limits on $m_{A^{'}}$ and $\epsilon$ under which our constraints are already ruled out. Such models therefore are unable to produce a RBG capable of explaining the EDGES results.

\section*{Acknowledgements}
The authors would like to thank Martin Winkler, Tim Linden and an anonymous referee for insightful comments on this work. RG and SZ acknowledges support by the Israel Science Foundation (grant no. 255/18). GM is supported by Swedish Research Council grant 2020-04691. 

\bibliographystyle{JHEP}
\bibliography{bibfile}

\end{document}